# Irradiation-induced broadening of the Raman spectra in monolayer graphene

I. Shlimak, A. Butenko, E. Kogan and M. Kaveh

*Jack and Pearl Resnick Institute of Advanced Technology, Department of Physics, Bar Ilan University, Ramat Gan 52900, Israel*

Abstract

Broadening of the Raman scattering (RS) spectra was studied in monolayer graphene samples irradiated with various dose of ions followed by annealing of radiation damage at different temperatures. It is shown that the width $\Gamma$ (full width at half maximum, FWHM) of three main RS lines (G-, D-, and 2D) increases linearly with increase of the density of irradiation-induced point defects $N_d$ as $\Delta\Gamma = m\, N_d$. The slope $m$ of the linear dependencies is the same for one-phonon emitting G-line and D-line, and almost double for two-phonon emitting 2D-line. It is also shown that the width of D-line $\Gamma_D$ for all samples is larger than one half of the width of 2D-line $\Gamma_{2D}$, which shows that in the case of D-line, elastic electron scattering on point defects leads to an additional decreasing the lifetime of the emitted phonon. Theoretical model of the width of D-line in disordered graphene is developed which explains the experimental observations and allows to determine the numerical coefficient in the in-plane transverse optic phonon dispersion in graphene.

## Introduction

Graphene is one atom thick planar sheet of $sp^2$-bonded carbon atoms arranged in a hexagonal lattice. The unit cell consists of two atoms which leads to separation of the graphene lattice into two nonequivalent sublattices [1]. Monolayer graphene is a two-dimensional gapless semiconductor where the conic valence and conduction bands contact each other in six single points (three K and three K') at the border of the Brillouin zone. Measurements of the Raman scattering (RS) spectra are widely used for study the quality of the graphene samples, characterization of defects and dopants (see, for example, [2-7]).

In the RS process, an absorbed photon generates an electron-hole pair and the excited electron with momentum $k$ is further scattered by phonons with momentum $q$ following by recombination with hole and emission of a photon red-shifted by the amount of energy given to the phonon. Due to momentum conservation requirements, the first-order RS process generates only phonons with wave vector $\boldsymbol{q} = 0$ at the $\Gamma$ point in the center of the first Brillouin zone. This band in RS spectra is known as G-band (~ 1580 cm$^{-1}$) and exists for all $sp^2$ carbon system, including amorphous carbon, carbon nanotubes and graphite.

In disordered graphene, when the lattice periodicity is broken near defects, the momentum conservation requirement is also broken. At first sight, in this case all phonons in the Brillouin zone are allowed for the RS, but the energy conservation requirement allocates in the resonant electron-phonon scattering some selected phonon modes in the vicinity of the K and K' points, which form, correspondingly, more intensive D-band (~ 1350 cm$^{-1}$, inter-valley scattering) with emission of in-plane transverse optic phonon (iTO) and less intense D'-band (~ 1620 cm$^{-1}$, intra-valley scattering) with emission of in-plane longitudinal optic phonon (iLO) [3].

The second-order RS process corresponds to excitation of two phonons with opposite momenta $q$ and $-q$ and can be observed in purely crystalline pristine samples since momentum is conserved. This gives rise to the intense 2D-line in monolayer graphene (~ 2700 cm$^{-1}$) which is due to emission of two iTO-phonons with opposite momenta. Fig. 1 shows the first Brillouin zone for monolayer graphene and sketch for G-, D-, and 2D- transitions.

In high quality pristine graphene, D-line is absent or barely seen, while 2D-line has maximal intensity. With increasing concentration of defects $N_d$, intensity of 2D-line $I_{2D}$ continuously decreases [8,9]. By contrast, with increase of $N_d$, the intensity of D-line $I_D$ first increases up to a maximum value and then it begins to decrease. Non-monotonic behavior of $I_D$ is explained on the basis of a model [8], in which D-line is emitted from some area $A$ with a radius $r_A$, surrounding the defect, therefore $I_D$ first increases and achieves the maximal value when the mean distance between defects $L_d \approx (N_d)^{-1/2}$ becomes of order of $r_A$. Further increase the density of defects leads to overlapping $A$-areas and decrease of $I_D$.

Ion irradiation induced disorder is not quite stable: some changes in the intensity of RS line were observed in irradiated samples after the long-term ageing (about one year) [10]. It was shown also that high-temperature annealing of radiation damage in vacuum and in the forming gas leads to partial reconstruction of the intensity of RS lines and lattice structure [11].

It should be noted that while changes in the intensity of lines were analyzed analytically [8], changes in the width of lines are studied in less details [9].

**Experiment**

In this work we study the broadening Γ (full width at half maximum, FWHM) for three main RS lines (G-, D-, and 2D-) for series of monolayer graphene samples gradually disordered by irradiation with Xe$^+$ and C$^+$ ions [10-12]. Fig. 2 shows the shape of D-line and 2D-line for samples irradiated with different dose Φ of Xe$^+$ ions having energy 35 keV. Disorder, introduced by irradiation with different ions, can be unified by using the density of irradiation-induced defects $N_d = k\Phi$, which is proportional to $\Phi$, and numerical coefficient $k$ depends on the energy $E$ and mass $M$ of the ion. The value of $k$ reflects the average fraction of carbon vacancies in the graphene lattice per ion impact. For $C^+$ ions with $M = 12$ in atomic mass units (amu) and for $Xe^+$ ions ($M = 131$ amu) with $E = 35$ keV, $k = 0.06$-$0.08$ and $0.8$ correspondingly [10]. The changes in the intensities of both D- and 2D-lines $I_D$ and $I_{2D}$ with increase of $N_d$ in these samples were studied in [10-12]. It was shown that $I_{2D}$ continuously decreases, while $I_D$ first increases and then decreases, the maximal value is achieved at $N_d \approx 4 \times 10^{12}$ cm$^{-2}$, which corresponds to the mean distance between defects $L_d \approx (N_d)^{-1/2} \approx 5$ nm. According to the model [8], this gives an estimation of the radius $r_A$ around a defect, where emission of one phonon can be accompanied by an elastic scattering on defect to satisfy momentum conservation.

Fig. 3 (a) shows dependencies of $\Gamma$ on $N_d$ for main RS lines in samples irradiated with Xe ions. All dependencies are nearly linear in the form of $\Gamma = \Gamma_0 + m\,N_d$, where $\Gamma_0$ is FWHM in the pristine sample, coefficient $m$ [in cm] characterizes the rate of broadening with increase of $N_d$. For one-phonon emitting D- and G- Raman lines, the values of $\Gamma_G$ and $\Gamma_D$ are close each other, and broaden with the same rate ($m \approx 0.9 \times 10^{-12}$ cm), while $\Gamma_{2D}$ is larger and broaden with almost double rate ($m \approx 1.7 \times 10^{-12}$ cm). This fact together with double energy of the peak position ($\omega_{2D} = 2\omega_D$) reflects the widely used description of the 2D-line as the second order of the D feature. Fig. 3 (b) shows the dependences of $\Gamma_{2D}$ on $\Gamma_D$ for different series of samples irradiated with $Xe^+$ and $C^+$ ions, both for initial samples (after irradiation) and for samples subjected to subsequent annealing at different temperatures $T_a$. These dependences look as the straight line with a slope about 2. For the first glance, the width $\Gamma_{2D}$ of the 2D-line caused by emission of two identical iTO phonons has to be double the width $\Gamma_D$ of the D-line connected with emission of only one iTO phonon. Dependence $\Gamma_{2D} = 2\,\Gamma_D$ is shown in Fig. 3(b) as the upper dashed line. One can see that the real values of $\Gamma_{2D}$ are less, especially for initial samples. In other words, the width $\Gamma_D$ is larger than one half of the width of 2D-line. This point is illustrated by Fig. 4 where both D- and 2D-lines are plotted together, but the upper scale for 2D-line is twice compressed. We see that the peak positions of D-line $\omega_D$ and the half of peak positions of 2D-line ($\omega_{2D}/2$) coincide for all samples independently on the degree of disorder. This confirms the elastic character of scattering by structural defects, so the energy of the emitted light quanta is changed only due to generation of one or two identical iTO phonons. However, it is also seen that the width of D-line is larger than one half of 2D-line. This difference could be attributed to the peculiarity of the D band process which, by contrast with 2D band, involves also an elastic scattering of electrons by a defect which has to accompany the emission of phonon for momentum conservation. Interaction with structural defects decreases the lifetime of emitted phonon which correspondingly leads to an additional broadening of D-line.

**Theory**

In an ideal crystalline monolayer graphene, the shape of the one-phonon Raman line $I(\omega)$ could be described by a Lorentzian function [9,13]

$$I(\omega) \propto \frac{1}{[\omega - \omega(q_0)]^2 + [\Gamma/2]^2} \qquad (1)$$

where $\omega$ is the frequency, $\omega(\boldsymbol{q}_0)$ is the peak position of the RS line associated with emission of a phonon of the wave vector $\boldsymbol{q}_0$ and $\Gamma$ is the full width at the half maximum (FWHM) which is reversely proportional to the phonon lifetime $\tau$, $\Gamma \sim 1/\tau$ due to uncertainty principle $\Delta E \sim \hbar/\tau$, which gives an indeterminacy in the value of the phonon energy, as measured in the Raman spectrum. Therefore, the line width in Raman spectra provide information on phonon lifetimes.

In an ideal crystal, the region over which the spatial correlation function of the phonon extends is infinite. This leads to the usual plane wave phonon eigenstates and the $\boldsymbol{q} = 0$ momentum selection rule of first order Raman scattering. However, as the crystal is damaged by ion bombardment, the mode correlation functions become finite due to the induced defects. Thus, there is a relaxation of the $\boldsymbol{q} = 0$ selection rule and associated with this relaxation, a finite correlation length $L$. As shown in [14], the assumption of a Gaussian attenuation factor exp (-$2r/L^2$), where $L$ is the size of the correlation region, leads upon Fourier transformation to an average over $q$ with a similar weighting factor $\exp[-q^2L^2/4]$. In the case of finite size of the correlation regions in the damaged material, the Raman intensity $I(\omega)$, can be written as [14]

$$I(\omega) \propto \int \exp\left(\frac{-q^2 L^2}{4}\right) \frac{dq}{[\omega^2 - \omega(q)]^2 - [\Gamma_0/2]^2} \qquad (2)$$

where $q$ is expressed in units of the reciprocal lattice constant, $\Gamma_0$ is the width of the unperturbed Raman line shape and $L$ is a measure of the phonon coherence length $L$ which should also a good measure of the average distance between point defects $L_d \approx (N_d)^{-1/2}$ [15,16].

Let us analyze the shape of the D-band. In this case, the general expression (2) is modified in the form [9]

$$I_D(\omega) \propto \int dq \frac{\exp\left(\frac{-(q-q_0)^2 L^2}{4}\right)}{[\omega - \omega_{iTO}(q)]^2 + [\Gamma_0/2]^2} \qquad (3)$$

Here $q_0 = 0.42$ Å$^{-1}$ is measured from the K point in the Brillouin zone [9], the phonon dispersion is taken in the form [16]: $\omega_{iTO}(q) = \omega_K + Aq - Bq^2$, where we will use only the first two terms:

$$\omega_{iTO}(q) = \omega_K + Aq \qquad (4)$$

In the integral (3) we change integration variable from $q$ to $\omega' = \omega_{iTO}(q) - \omega_{iTO}(q_0)$ and also measure $\omega$ from $\omega_{iTO}(q_0)$. We obtain

$$\omega' = \omega_{iTO}(q) - \omega_{iTO}(q_0) \approx A(q - q_0) \qquad (5)$$

Thus, integral (3) can be approximated as

$$I_D(\omega) \propto \int G(\omega';\sigma) L((\omega - \omega'');\gamma) d\omega' \qquad (6)$$

where $\gamma = \Gamma_0/2$, $\sigma = (2)^{1/2} A/L$. In Eq. (6)

$$G(x;\sigma) \equiv \frac{e^{-x^2/(2\sigma^2)}}{\sigma\sqrt{2\pi}} \qquad (7)$$

is the centered Gaussian profile, and

$$L(x;\gamma) \equiv \frac{\gamma}{\pi(x^2 + \gamma^2)} \qquad (8)$$

is the centered Lorentzian profile.

The r.h.s. of Eq. (6) is called the Voigt distribution [17]. In spectroscopy, a Voigt profile results from the convolution of two broadening mechanisms, one of which alone would produce a Gaussian profile (usually, as a result of the Doppler broadening), and the other would produce a Lorentzian profile. Voigt profiles are common in many branches of spectroscopy and diffraction. The FWHM of the Voigt profile can be found from the associated Gaussian and Lorentzian widths [17]. The FWHM of the Gaussian profile is

$$f_G = 2\sigma\sqrt{2\ln(2)} \qquad (9)$$

The FWHM of the Lorentzian profile is

$$f_L = 2\gamma \qquad (10)$$

A rough approximation for the relation between the widths of the Voigt, Gaussian, and Lorentzian profiles is:

$$f_V \approx f_L / 2 + \sqrt{\left(f_L / 2\right)^2 + f_G^2} \qquad (11)$$

In our case

$$f_L = \Gamma_0, \qquad f_G = 4(\ln 2)^{1/2} A/L \qquad (12)$$

so Eq. (11) gives

$$f_V \approx \frac{\Gamma_0}{2} + \sqrt{\frac{\Gamma_0^2}{4} + \frac{16(\ln 2)A^2}{L^2}} \qquad (13)$$

In the limit $f_G \gg f_L$ which corresponds to the case of strong broadening ($\Delta\Gamma \gg \Gamma_0$), we obtain

$$f_V \approx \left. f_L \middle/ 2 \right. + f_G \approx \frac{\Gamma_0}{2} + \frac{4\sqrt{\ln 2}A}{L} \qquad (14)$$

So, in this case, the broadening of the Raman line $\Delta\Gamma$ is proportional to $1/L$ or to the concentration of defects to the power one half $(N_d)^{1/2}$: $\Delta\Gamma \sim 1/L \approx 1/L_d \sim (N_d)^{1/2}$.

To analyze the opposite limit $f_G \ll f_L$ (weak broadening $\Delta\Gamma \leq \Gamma_0$), it is convenient to present Eq. (13) as

$$f_V \approx \frac{\Gamma_0}{2}\left[1 + \sqrt{1 + \frac{64(\ln 2)A^2}{\Gamma_0^2 L^2}}\right] \qquad (15)$$

Using the relation $(1 + x)^{1/2} \approx 1 + x/2$, valid for $x \ll 1$, in the limit $f_G \ll f_L$, we obtain

$$f_V \approx f_L + \frac{f_G^2}{f_L} \approx \Gamma_0 + \frac{16(\ln 2)A^2}{\Gamma_0 L^2} \qquad (16)$$

so the broadening of the Raman line is proportional to the concentration of defects $N_d$: $\Delta\Gamma \sim (1/L)^2 \approx (1/L_d)^2 \sim N_d$.

### Discussion

The experiment (Fig. 3a) clearly shows that the broadening of the D-line is linearly proportional to the density of irradiation-induced defects $\Delta\Gamma \sim N_d$. This indicates that Eq. (16) correctly describes the broadening which can be explained by the fact that the broadening induced by disorder and associated with the Gaussian profile $f_G$ is not much larger than the initial line width $\Gamma_0$ associated with the Lorentzian profile. Indeed, we see in Fig. 3a that the width of D-line for pristine sample $\Gamma_0 = 30$ cm$^{-1}$, while for highly disordered sample with $N_d = 3.5 \times 10^{13}$ cm$^{-2}$, $\Gamma = 60$ cm$^{-1}$, so $\Delta\Gamma \approx \Gamma_0$.

Eq. (16) allows us to determine the value of numerical coefficient $A$ in the expression for iTO phonon dispersion, Eq. (4). From comparison of experimental dependence $\Gamma = \Gamma_0 + mN_d$ and theoretical expression (16) we get $m = 16(\ln2) \, A^2/\Gamma_0$. Using $m = 0.9 \times 10^{-12}$ cm and $\Gamma_0 \approx 30$ cm$^{-1}$ (see Fig. 3a), we obtain $A = 1.55 \times 10^{-6}$ if $q$ is measured in cm$^{-1}$ or, correspondingly, $A = 155$, if $q$ is measured in reciprocal Angstroms (Å$^{-1}$)

The similar linear dependence of $\Gamma_G$ and $\Gamma_D$ as a function of $N_d$ shown in Fig. 3a implies that there is a common mechanism of the additional decrease of the phonon lifetime connected with irradiation-induced structural defect. According to [18], three phenomena must be taken into account in estimating the scattering of phonons by point defects: the change in mass due to the presence of the defect, the change in the interatomic bonds close to the defect and the elastic strain field around the defect. It was shown in [19], that in strained pristine sample, the broadening leads to a linear dependence between $\Gamma_{2D}$ and $\Gamma_G$ with a slope $s = 2.2$. In our case, $\Gamma_{2D}$ is, indeed, linearly dependent on $\Gamma_G$, but with smaller slope 1.4 (Fig. 5). This means that strain is important but not the only reason for the broadening of Raman lines.

It is seen in Fig. 3 (b) that after annealing, the width $\Gamma_{2D}$ becomes closer to $2\,\Gamma_D$ which is shown as the upper dashed line. The same effect can also be seen in Fig. 4 (d) for the sample with high density of defects. It means that the influence of an additional decrease of $\tau$ due to point defects becomes less important, which can be explained by an enhancement of the general disorder of the graphene film. For example, it was shown in [11] that after high temperature annealing, the strained areas are located not only near the point defects, but are spread across the whole film. The propagation of phonon scattering phenomena throughout the whole film in strongly disordered samples should also lead to a decrease in the effect of point defects.

In summary, the broadening of three main RS lines (G-, D-, and 2D-) has been studied in monolayer graphene samples gradually disordered by ion irradiation. It is shown experimentally that FWHM for all lines increases linearly with increase of the density of irradiation-induced point defects. Theoretical model of the width of D-line in disordered graphene is developed which explain the experimental observations and reveal to determine the numerical coefficient in the expression for in-plane transverse optic phonon dispersion in graphene.

**Acknowledgements**

We are thankful to E. Zion for his help in Raman measurements and to the Erick and Sheila Samson Institute of Advanced Technology for support.

Figure captions

Fig. 1. (a) - first Brillouin zone in monolayer graphene, (b-d) sketch of the Raman scattering process for G-, D- and 2D-lines correspondingly.

Fig. 2. The shape of D-line and 2D-lines for samples irradiated with different dose of Xe ions with energy 35 keV. Dose of irradiation $\Phi$ (in units of $10^{13}$ cm$^{-2}$): 0 - 0 (initial), 1 – 0.15, 2 – 0.3, 3 – 0.5, 4 – 1.0, 5 – 2.0, 6 – 4.0. Intensity of lines are normalized with respect to the intensity of G-line. Dotted lines show the Lorentzian fit.

Fig. 3. (a) - dependences of $\Gamma_D$, $\Gamma_G$ and $\Gamma_{2D}$ on the density of point defects $N_d$ for samples from Xe$^+$-series; (b) - dependences $\Gamma_{2D}$ on $\Gamma_D$ for initial samples and samples from C$^+$-series subsequently annealed at different temperatures $T_a$. The upper dashed line corresponds to the relation $\Gamma_{2D} = 2\,\Gamma_D$.

Fig. 4. D-line (1), lower scale, and 2D-line (2), upper twice suppressed scale, for some samples from Xe-series. $N_d$ (in units of $10^{13}$ cm$^{-2}$): a – 0.12, b – 0.24, c – 0.4, d – 0.8.

Fig. 5. Dependences of $\Gamma_D$ (1) and $\Gamma_{2D}$ (2) on $\Gamma_G$ for samples from Xe-series. The slope $s$ of the linear dependence is shown near the line.

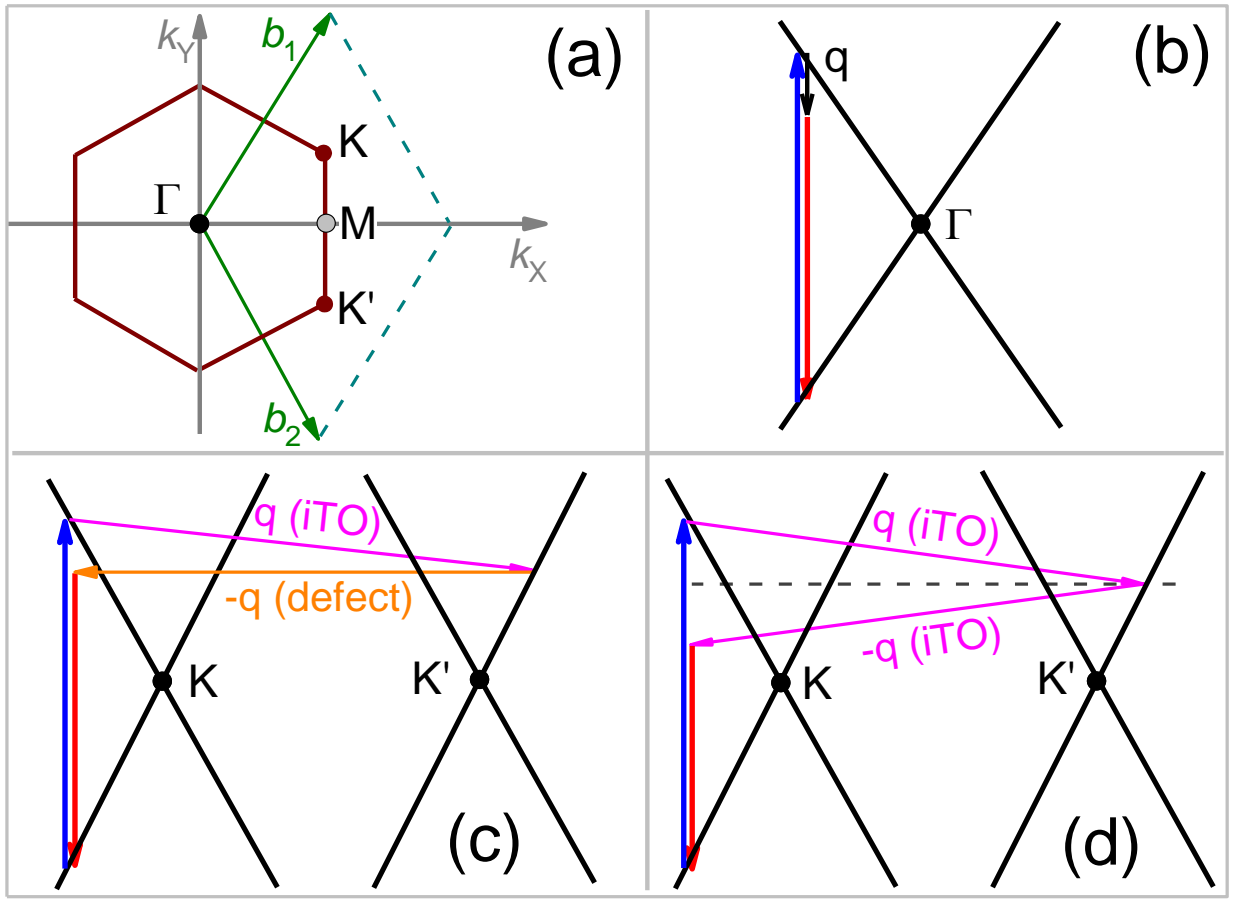

Fig. 1

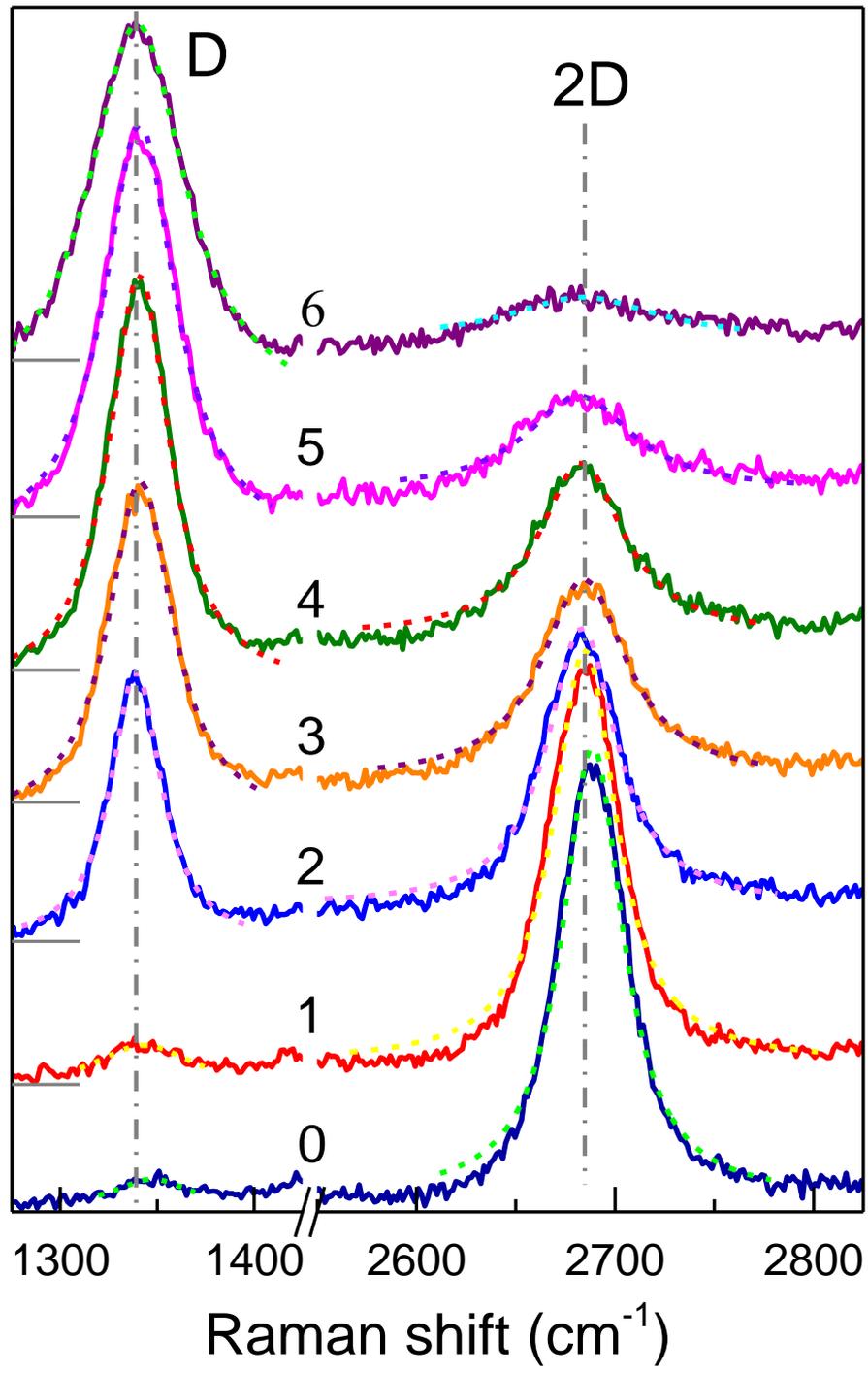

Fig. 2

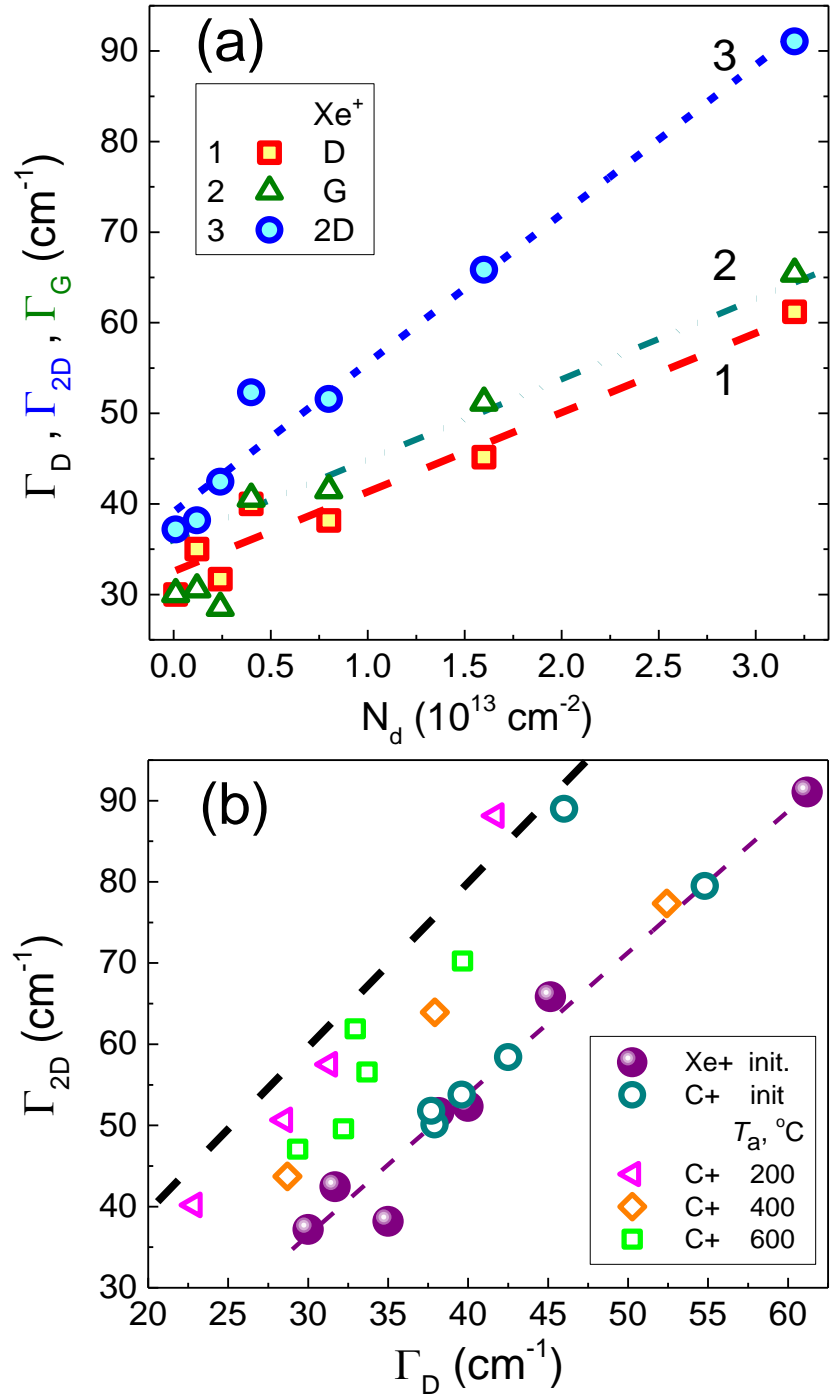

Fig. 3

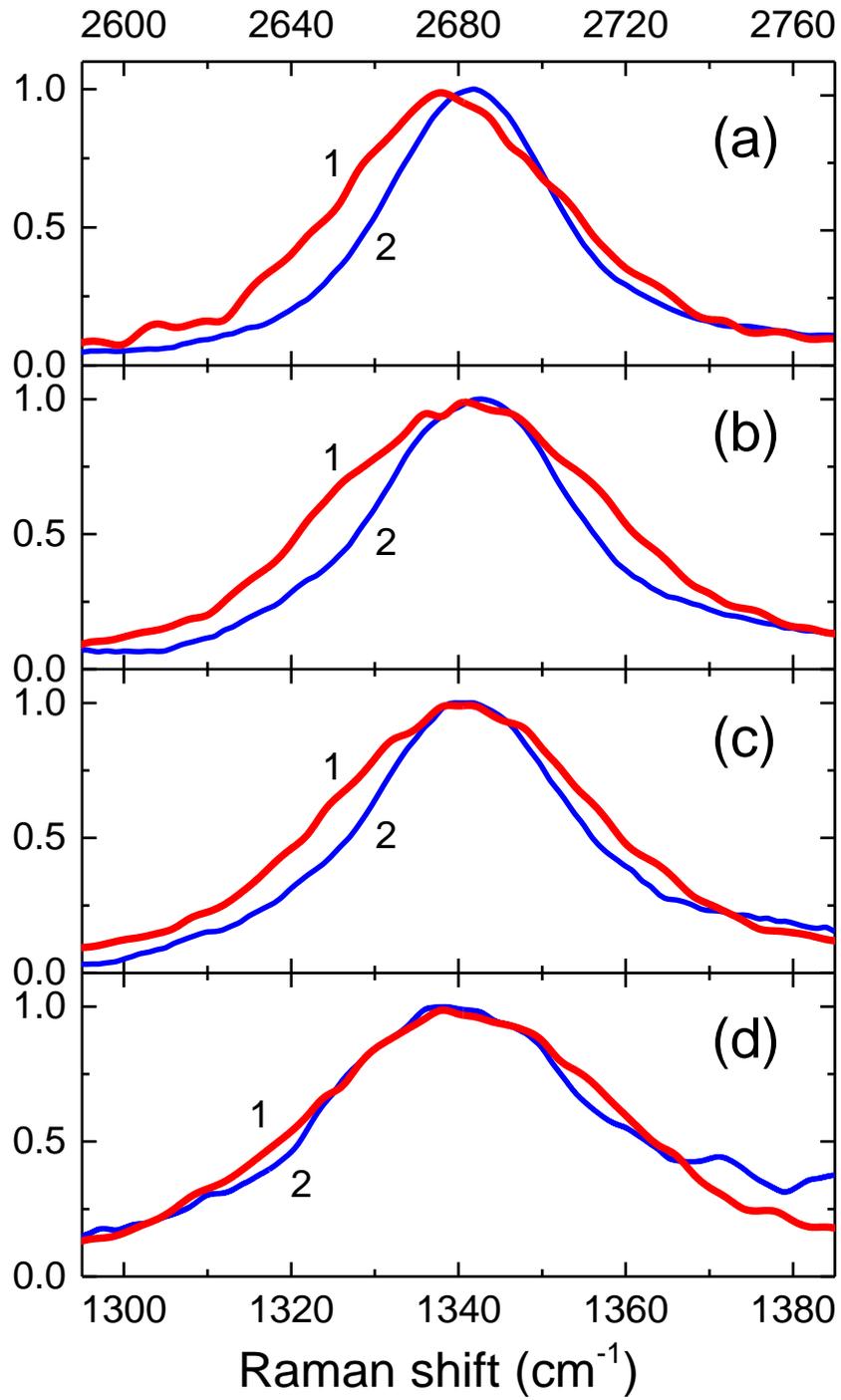

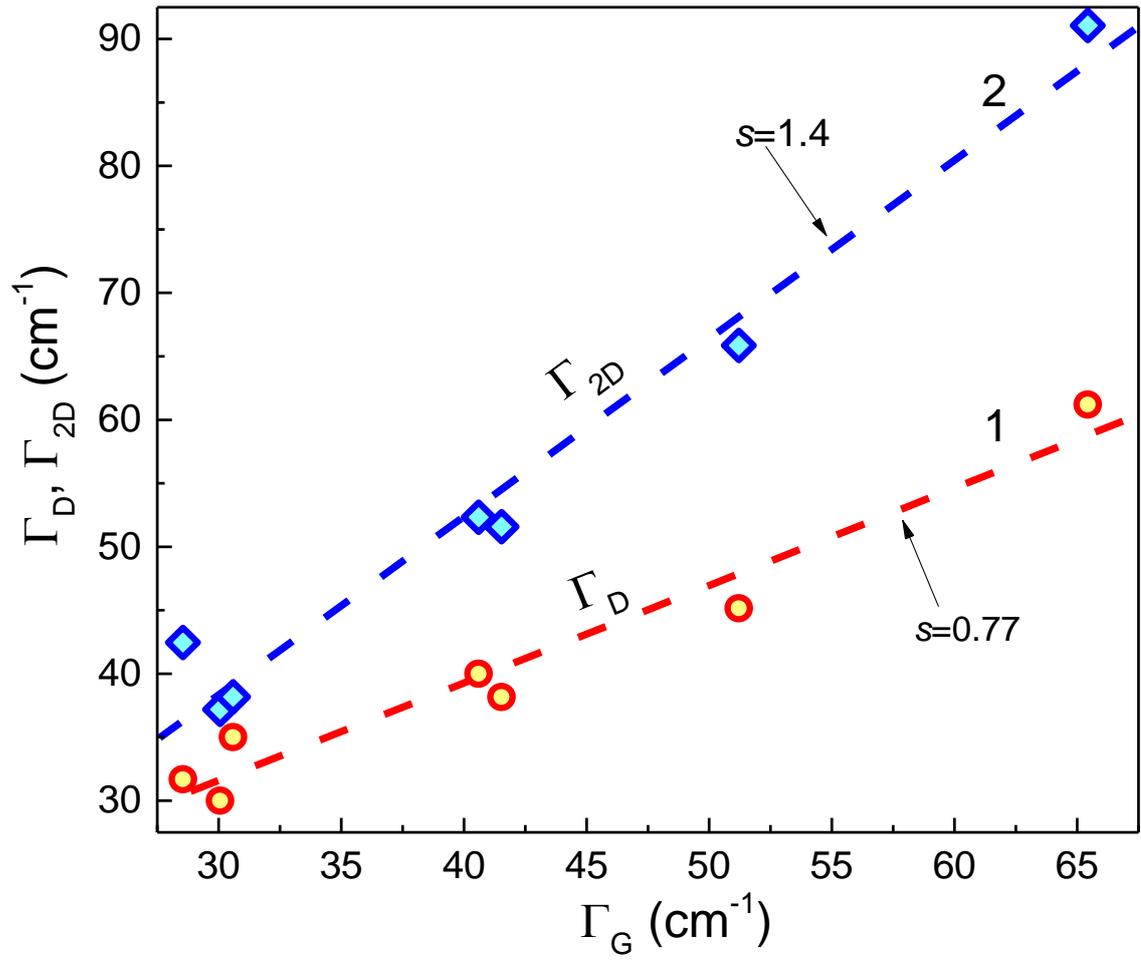

Fig. 5